\documentclass{amsart}
\usepackage{amssymb}
\usepackage{amsfonts}

\theoremstyle{plain}

\newtheorem{lemma}{Lemma}

\newtheorem{proposition}{Proposition}

\numberwithin{equation}{section}
\input{tcilatex}

\begin{document}
\title[Exotic option pricing]{A comprehensive method for exotic option
pricing}
\address{Department MatemateS, University of Bologna\\
viale Filopanti n.5 - Bologna (Italy)\\
and Faculty of Economics in Rimini, via Angher\`{a} n.22 - Rimini (Italy)}
\email{rossella.agliardi@unibo.it}
\date{October 12, 2009}

\begin{abstract}
This work illustrates how several new pricing formulas for exotic options
can be derived within a L\`{e}vy framework by employing a unique pricing
expression. Many existing pricing formulas of the traditional Gaussian model
are obtained as a by-product.
\end{abstract}

\author{Rossella Agliardi}
\maketitle

\section{Introduction}

In the last decades the literature on option pricing under L\'{e}vy
processes mushroomed. The reason is that L\'{e}vy processes have the
flexibility to capture empirical characteristics in stock returns such as
jumps in price due to market shocks and distributional properties such as
skewness and semi-heavy tails, while maintaining most of the analytical
tractability of the Gaussian model. In this paper we adopt the class of
regular L\'{e}vy processes of exponential type (RLPE) as the driving
processes, following [7]. As [7] points out, it is the most tractable
subclass of L\'{e}vy process from the analytical point of view if the
Brownian motion is not available. Their characteristic exponents $\psi (\xi
) $ enjoy very favourable properties as symbols of pseudo differential
operators, since their real part behaves as $c\left\vert \xi \right\vert
^{\upsilon }$ \ as $\left\vert \xi \right\vert \rightarrow \infty $ in the
strip of regularity, with positive $c$ and $\upsilon $. Thus the integrals
appearing in the pricing formulas are absolutely convergent thanks to the
terms of the form $e^{-\tau \psi (\xi )}$. Moreover, one can differentiate
under the integral sign or shift the line of integration by using the Cauchy
theorem for holomorphic functions. Such properties allow for a great
flexibility of the method when working out the analytical pricing formulas
for several exotic options. Moreover, such a class incorporates most of the
models which have been proposed as an alternative to the traditional
geometric Brownian motion, including the Normal Inverse Gaussian (NIG)
(Barndorff-Nielsen [6]), the Hyperbolic (H) (Eberlein and Keller [13]) and
the more general Generalized Hyperbolic (GH) (Eberlein and Prause [15]), the
four-parameter distribution named CGMY after the names of Carr, Geman, Madan
and Yor [10] and generalized in [11], just to quote the most popular ones.
The development of pricing models replacing the traditional underlying
source of randomness, the Brownian motion, by a L\'{e}vy process has
fostered a good deal of work on exotic option pricing which parallels the
existing results in the Gaussian framework. (See [27], for a compendium of
recent research on the topic).

In this article the focus is on the pricing of European exotics and the aim
is to present a valuation formula which is the most comprehensive as
possible, in that several types of options can be priced directly and no
specific method has to be devised for each of them. The formula is tailored
to valuate discretely monitored options, which are the most popular ones in
view of the regulatory issues and the trading practice. However, the
continuous counterpart can be derived in some cases (see Example 2). The
method employed in this work is to start from (multi-period) digital options
as building-blocks and then to price the more complex options in terms of
such elementary contracts. The idea that a broad class of financial
derivatives can be evaluated in terms of elementary contracts such as
digital options has been applied in the traditional Gaussian framework (see
[9] and [23]) but it has not yet fully exploited in a non-Gaussian modeling.
By providing the non-Gaussian counterpart of this view, we are able to
obtain new pricing formulas in the L\`{e}vy environment and to throw a new
insight into some known pricing expressions. This method has been
anticipated in [3]. This work employs a more general setting and provides
further examples of exotic option prices. Since a Brownian motion is a RLPE
of order 2 and any exponential type - and thus is captured in this framework
- each example includes the known pricing expressions of the classical
Gaussian modeling. An emphasis is put on deriving the existing formulas of
the traditional Black-Scholes modeling from our more general framework, so
that the reader is confronted with the more familiar expressions. To our
surprise, the Lemma which is employed to the purpose is not found in the L%
\'{e}vy literature and therefore a proof \ is given. Some applications of
the main result are given in Section 4. They serve as an illustration of the
flexibility of our valuation formula and integrate the several examples
developed by the author in a previous work [3].

This paper is organized as follows. Section 2 outlines the main definitions
and notation concerning L\'{e}vy processes and provides a Lemma linking the
Gaussian multivariate distribution to the Fourier transform-based L\`{e}vy
setting. This Lemma will be used extensively throughout the paper and has a 
\textit{per se} mathematical interest. Section 3 presents the general
valuation formula, while some examples of exotic option prices are provided
in Section 4. The choice of the examples has been dictated by the wish to
provide new valuation formulas and presenting them in a general form (for
example, N-fold compound options instead of just 2-fold ones,\ flexible
Asian options, both call and put options in one expression, etc.). However
they are by no means exhaustive with respect to the potentialities of the
approach.\bigskip

\section{Preliminaries}

Let us consider a L\'{e}vy market, i.e. a model of financial market with a
deterministic saving account $e^{rt}$, $r\geq 0$, and a stock following a
stochastic process $S_{t}=e^{X_{t}}$, where $(X_{t})_{t\geq 0}$ is a L\'{e}%
vy process. As usual in a L\'{e}vy setting, the L\'{e}vy process replaces
the Brownian motion which is employed in the classical modeling of stock
prices. Here we assume that the stock price $S_{t}$ is $e^{X_{t}}$, where $%
(X_{t})_{t\geq 0}$ is a one-dimensional RLPE, i.e. a regular L\'{e}vy
process of order $\upsilon \in ]0,2]$ and exponential type $[\lambda
_{-},\lambda _{+}]$ , $\lambda _{-}<0<\lambda _{+}$. This means that $%
(X_{t})_{t\geq 0}$\ has a characteristic exponent $\psi (\xi )$ which admits
a representation of the form:\newline
\begin{equation}
\psi (\xi )=-i\mu \xi +\phi (\xi )
\end{equation}%
\newline
where $\phi $ is holomorphic in the strip $\func{Im}\xi \in ]\lambda
_{-},\lambda _{+}[$, continuous up to the boundary of the strip, and $\phi
(\xi )=C\left\vert \xi \right\vert ^{\upsilon }+O(\left\vert \xi \right\vert
^{\upsilon _{1}})$ for $\left\vert \xi \right\vert \longrightarrow \infty $
\ and $\left\vert \phi ^{\prime }(\xi )\right\vert \leq C(1+\left\vert \xi
\right\vert ^{\upsilon _{2}})$ \ in $\func{Im}\xi \in \lbrack \lambda
_{-},\lambda _{+}]$ \ with $\upsilon _{1},\upsilon _{2}<\upsilon .$ (See
[B-L] for a comprehensive and Finance-oriented theory of RLPE processes). In
order to price contingent claims on the stock, it is convenient to consider
an equivalent martingale measure (EMM) $Q$ which makes the discounted price
process $e^{-rt}S_{t}$ a martingale. Let $\psi _{P}$ (respectively $\psi
_{Q} $) denote the characteristic exponent with respect to the historic
measure $P $ (an EMM $Q$, respectively), i.e. $E_{P}(e^{i\xi
X_{t}})=e^{-t\psi _{P}(\xi )}$ \ ($E_{Q}(e^{i\xi X_{t}})=e^{-t\psi _{Q}(\xi
)}$, respectively). We assume that the discounted stock price $e^{-rt}S_{t}$
is a martingale under $Q$, that is, we assume that the equivalent martingale
measure condition (EMM-condition) \ $r+\psi _{Q}(-i)=0$ holds. Furthermore,
the additional condition $\lambda _{-}<-1$, which is usually assumed to let $%
-i$ belong to the strip of regularity of $\psi $, is supposed to hold true.
In the sequel further restrictions on $\lambda _{\pm }$ will be posed,
depending on the kind of exotic option under study. We recall that $Q$ is
not unique in a general L\'{e}vy setting. We do not dive into the problem of
choosing a martingale measure and refer to the well-established literature
on L\'{e}vy processes in Finance dealing with this topic extensively (see
[17], for example). A popular method is based on Esscher transform and the
expression for $\psi _{Q}(\xi )$ is obtained by first solving the equation \ 
$\psi _{P}(-ih)-\psi _{P}(-ih-i)=r$ for $h$, and then letting \ $\psi
_{Q}(\xi )=\psi _{P}(\xi -ih)-\psi _{P}(-ih).$ In what follows we assume
that an EMM $Q$ is chosen so that $X_{t}$ is a RLPE under it and we will
omit the subscript $Q$ both in $E$ and $\psi $.

Let $g(X_{T})$ denote the terminal payoff of an option on $S_{t}$ at the
expiry date $T$. Then the no-arbitrage price of the option at the current
time $t$ ($t<T$) is given by:

\begin{equation}
F(S_{t},t)=E[e^{-r(T-t)}g(X_{T})\mid X_{t}=\ln (S_{t})]
\end{equation}%
An explicit expression for $F(S_{t},t)$ can be obtained in terms of the
Fourier transform $\widehat{g}(\xi )$ in the complex plane. Indeed, if $%
e^{\omega x}g(x)\in L^{1}(%
\mathbb{R}
)$ \ for some $\omega \in ]\lambda _{-},\lambda _{+}[$ , then the Fourier
transform $\widehat{g}(\xi )$ of $g$ can be defined on $\func{Im}\xi =\omega 
$ and one obtains:\newline
\begin{equation}
F(S_{t},t)=\frac{1}{2\pi }\int\nolimits_{-\infty +i\omega }^{+\infty
+i\omega }e^{i\xi \ln S_{t}-(T-t)(r+\psi (\xi ))}\widehat{g}(\xi )d\xi
\end{equation}

\noindent In the following sections this set-up is extended to the more
general case of path-dependent options where the payoff $g$ depends on a set
of fixed asset price monitoring times, $T_{1}<...<T_{M}\leq T$, so that
several types of exotic options can be priced in this framework.

Finally we recall that a Brownian motion is a RLPE of order 2 and any
exponential type. Therefore the pricing formulas of the traditional Gaussian
approach can be obtained in our framework for several exotics. To the
purpose we need the following:

\begin{lemma}
Let $C$ $=(c_{kj})$ denote an $N\times N$ correlation matrix. Then for any $%
\omega _{k}>0$, for any real number $d_{k}$ \ and with $w_{k}=\pm 1$, $%
k=1,...N$, the following identity holds:\medskip \newline
$\frac{1}{(2\pi i)^{N}}\int\nolimits_{-\infty -iw_{N}\omega _{N}}^{+\infty
-iw_{N}\omega _{N}}$ ...$\int\nolimits_{-\infty -iw_{1}\omega _{1}}^{+\infty
-iw_{1}\omega _{1}}e^{\Sigma _{k=1}^{N}i\xi _{k}d_{k}-\frac{1}{2}\Sigma
_{k,j=1}^{N}c_{kj}\xi _{k}\xi _{j}}\frac{1}{\Pi _{k=1}^{N}\xi _{k}}d\xi
_{1}...d\xi _{N}$ $=\medskip $\newline
$=[\tprod\limits_{k=1,...,N}w_{k}]N_{N}(w_{1}d_{1},...w_{N}d_{N};WCW)%
\medskip $\newline
where $N_{N}$ denotes the $N$-variate multinormal cumulative distribution
function and $W$ is the $N\times N$ diagonal matrix with entries $w_{k}$, $%
k=1,...N$.$\medskip $
\end{lemma}

\noindent \textbf{Proof.} \ Consider the case $w_{k}=1$, $k=1,...N$, at
first. Let $\Omega =(\omega _{1},...,\omega _{N})$ and let $\varkappa
_{\Omega }(x_{1},...,x_{N})$ denote $\exp (-\tsum\limits_{j=1,...,N}\omega
_{j}x_{j})I_{[-d_{1},+\infty )}(x_{1})...I_{[-d_{N},+\infty )}(x_{N})$ \
which belongs to $L^{1}(%
\mathbb{R}
^{N})$. Its Fourier transform in $(\xi _{1}^{R},..,\xi _{N}^{R})\in 
\mathbb{R}
^{N}$ is:\newline
$\exp [i\tsum\limits_{j=1,...,N}d_{j}\xi _{j}]/\tprod\limits_{j=1,...,N}i\xi
_{j}$ , \ \ with $\xi _{j}=\xi _{j}^{R}-i\omega _{j}$. On the other hand, $%
\exp (-\frac{1}{2}\Xi ^{T}C\Xi )$ is the Fourier transform of $\phi (X)=%
\frac{1}{(2\pi )^{M/2}\sqrt{\det C}}\exp (-\frac{1}{2}X^{T}C^{-1}X).$ Thus
the left-hand side term can be written as a convolution of $\varkappa
_{\Omega }$ and $\phi $, that is, as\newline
$\frac{1}{(2\pi )^{M/2}\sqrt{\det C}}\int_{%
\mathbb{R}
^{N}}\varkappa _{\Omega }(-Y)e^{-\Omega ^{T}Y}\exp (-\frac{1}{2}%
Y^{T}C^{-1}Y)dY$,\newline
which is \ $\int\nolimits_{-\infty }^{d_{1}}...\int\nolimits_{-\infty
}^{d_{N}}\frac{\exp (-\frac{1}{2}Y^{T}C^{-1}Y)}{(2\pi )^{M/2}\sqrt{\det C}}%
dy_{1}...dy_{N}$ $=$ $N_{N}(d_{1},...d_{N};C).$ The general case follows
changing to variables $\xi _{j}^{\prime }=w_{j}\xi _{j}$ in the integrals of
the left-hand term and replacing each $d_{k}$ with $\widehat{d}_{k}$ $%
=w_{k}h_{k}$ and $C$ with $WCW$ in the previous argument.\bigskip

\section{Set-up and main result}

In this section an almost universal option pricing formula is proved within
the L\'{e}vy framework described in the previous section. In particular we
derive the arbitrage-free price for generalized multi-period exotic power
digital options. Such options are the building blocks for a broad class of
exotic options with a single underlying asset, because several exotic
options can be expressed as static portfolios of these multi-period power
digitals. The idea of pricing several exotic options by a single universal
formula is due to [9], where the classical Gaussian case was studied. Thus
this Section provides the generalization to the L\'{e}vy framework and the
resulting formula is a most comprehensive one, both in terms of the kind of
option and of the stochastic process driving the underlying asset. Since the
aim is to establish a unifying set-up, the formal notation is a bit involved
and is laid down along the lines of [9].

Assume that the payoff of an option depends on $M$ fixed asset price
monitoring times, $T_{1}<...<T_{M}$, where, for simplicity's sake, $T_{M}=T$%
, the expiry date of the option. Let \textbf{T}\textit{\ }denote the set of
times $[t,T_{1},...,T_{M}]$, where $t$ is the current time, $t<T_{1}.$ Let $%
S_{k}$ denote the price of the underlying asset at the monitoring time $%
T_{k} $ and let \textbf{S }denote the $M$-dimensional vector assembling all
the components $S_{k}$ which are relevant for the option under study. If $%
\gamma $ is $M$-dimensional , then \textbf{S}$^{\gamma }$ denotes $%
S_{1}^{\gamma _{1}}...S_{M}^{\gamma _{M}}$ and $\gamma $ is referred to as
the payoff index vector. We assume that $S_{t}=e^{X_{t}}$, where $X_{t}$ is
a L\'{e}vy process. Thus, in most cases, it will be convenient to work
directly with $X_{k}$, the value of $X$ at $T_{k}$, and with the vector 
\textbf{X}$=(X_{1},...,X_{M})$. Since we want to treat call and put options
together, we introduce $W$, a diagonal matrix with all the diagonal entries $%
w_{ii}$ equal to $\pm 1$. To simplify notation, $w_{ii}$ will be denoted by $%
w_{i}$. Then both the indicator functions $I_{[K_{i},+\infty )}(X_{i})$ and $%
I_{(-\infty ,K_{i}]}(X_{i})$ may be encompassed in a unique notation, $%
\mathbf{1}_{1}(w_{i}X_{i}\geq w_{i}K_{i})$, depending on the sign of $w_{i}$%
. Let \textbf{K} denote the exercise price vector (of dimension $N$) and let 
\textbf{1}$_{N}(\mathbf{Y}\geq \mathbf{K})$ denote the $N-$dimensional
indicator function \ $\prod\limits_{i=1}^{N}\mathbf{1}_{1}(Y_{i}\geq K_{i}).$
In order to give a greater flexibility to the approach, the exercise
condition matrix is introduced. Such term will denote any $N\times M$ matrix 
$A=(a_{nk})$ involved in the payoff, where $N$ is the exercise dimension.
For example, the payoff of discrete mean Asian options depends on \ $\sqrt[M]%
{\Pi _{k=1,...,M}\ S_{k}}\lessgtr \widetilde{K}$ or, equivalently, $%
\tsum\limits_{k=1,...M}\frac{1}{M}X_{k}\lessgtr \ln \widetilde{K}=K.$ Thus
their payoff can be expressed in terms of \ $\mathbf{1}_{1}(w_{1}A$\textbf{X}%
$\geq w_{1}K)$, where $A=[\frac{1}{M},...,\frac{1}{M}]$ is a $1\times M$
matrix. On the other hand, the payoff of a discretely monitored digital
option \textbf{1}$_{M}(WX\geq WK)$ can be expressed in the general form 
\textbf{1}$_{M}(WA\mathbf{X}\geq W\mathbf{K})$ by taking $A$ as the $M\times
M$ identity matrix. All the useful parameters are summarized in the payoff
parameter set \textbf{P}$=[(\gamma _{1},...\gamma _{M}),\mathbf{K},W,A].$

Let $F(S_{t},t;$\textbf{T}$\mathbf{,}$\textbf{P}$)$ denote the value of a
multi-period binary whose payoff is specified in terms of \textbf{T }and%
\textbf{\ P }and where $S_{t}$ denotes the value of the underlying asset%
\textbf{\ }at the current time $t$. More specifically, the expiry $T$ payoff
function is \ $\exp [\tsum\limits_{k=1}^{M}\gamma _{k}X_{k}].$\textbf{1}$%
_{N}(WA\mathbf{X}\geq W\mathbf{K}).$ In most applications $\gamma _{k}=0$
for $k\neq M.$ The following Proposition gives the arbitrage-free current
value of this generalized multi-period option under the assumption that the
value $S_{t}$ of the underlying asset $S_{t}$ is $e^{X_{t}}$, where $X_{t}$
is a RLPE of order $\upsilon \in ]0,2]$ and exponential type $[\lambda
_{-},\lambda _{+}].$ The main result of the paper is the following:

\begin{proposition}
The following valuation formula holds:\newline
$F(S_{t},t;$\textbf{T}$\mathbf{,}$\textbf{P}$)=\frac{e^{-r(T_{M}-t)}}{(2\pi
i)^{N}}S_{t}^{\tsum\limits_{k=1}^{M}\gamma
_{k}}\tprod\limits_{n=1}^{N}w_{n}\int\nolimits_{-\infty -iw_{N}\omega
_{N}}^{+\infty -iw_{N}\omega _{N}}...\int\nolimits_{-\infty -iw_{1}\omega
_{1}}^{+\infty -iw_{1}\omega _{1}}{}^{\text{ }}\frac{1}{\tprod%
\limits_{n=1,...N}\xi _{n}}$\newline
$\exp [i\tsum\limits_{n=1}^{N}\xi _{n}(\tsum\limits_{k=1}^{M}a_{nk}\ln
S_{t}-K_{n})-\Psi (t,\xi _{1},...\xi _{N})]d\xi _{1}..d\xi _{N}$\newline
where $\Psi (t,\xi _{1},...\xi
_{N})=\tsum\limits_{j=1}^{M}(T_{j}-T_{j-1})\psi
(\tsum\limits_{n=1}^{N}\tsum\limits_{k=j}^{M}a_{nk}\xi
_{n}-i\tsum\limits_{k=j}^{M}\gamma _{k})$ \ with $T_{0}=t$, $\omega _{n}>0$
for $n=1,...N$ \ and $\tsum\limits_{k=j}^{M}\left[ \tsum%
\limits_{n=1}^{N}w_{n}\omega _{n}a_{nk}+\gamma _{k}\right] \in ]-\lambda
_{+},-\lambda _{-}[$ \ for $j=1,...,M.\bigskip $
\end{proposition}

\noindent Before proving Proposition 1 we give a pricing formula for the
simple case of a power digital option, which\ is slightly more general than
Proposition 1 in [3].

\begin{lemma}
Let $F(S_{t},t)$ denote the current arbitrage-free price of the power option
with expiry $T$ payoff function $S_{T}^{\gamma }\mathbf{1}_{1}(waX_{T}\geq
wK)$, $a\neq 0$, $w=\pm 1$, $X_{t}=\ln S_{t}.$ Then for any $\omega >0$ such
that $aw\omega +\gamma \in ]-\lambda _{+},-\lambda _{-}[$ ,\ one can write%
\newline
\begin{equation}
F(S_{t},t)=\frac{wS_{t}^{\gamma }}{2\pi i}\int\nolimits_{-\infty -iw\omega
}^{+\infty -iw\omega }e^{i\xi \lbrack a\ln S_{t}-K]-(T-t)[r+\psi (a\xi
-i\gamma )]}\frac{1}{\xi }d\xi
\end{equation}
\end{lemma}

\noindent \textbf{Proof}. The Fourier transform of the payoff function $%
g(X_{T})=e^{\gamma X}$ $\mathbf{1}_{1}(waX_{T}\geq wK)$ is $\ \widehat{g}%
(\eta )=\frac{w.sgn(a)}{i(\eta +i\gamma )}\exp [(\gamma -i\eta )\frac{K}{a}]$
\ \ \ with $\func{Im}\eta =-w\omega _{\gamma }$ \ for any $\omega _{\gamma }$
such that $\gamma -w\omega _{\gamma }\gtrless 0$ whenever $aw\lessgtr 0$ and 
$w\omega _{\gamma }\in ]-\lambda _{+},-\lambda _{-}[.$ Then:\medskip \newline
$F(S_{t},t)=\frac{w.sgn(a)}{2\pi i}\int\nolimits_{-\infty -iw\omega _{\gamma
}}^{+\infty -iw\omega _{\gamma }}e^{i\eta \lbrack \ln (S_{t})-\frac{K}{a}%
]-(T-t)(r+\psi (\eta ))}\frac{1}{\eta +i\gamma }d\eta \medskip $\newline
which gives (3.1) changing to variables $\eta +i\gamma =a\xi $ and letting $%
\omega _{\gamma }-w\gamma =a\omega $.\bigskip

\noindent \textbf{Remark 1}. The expression (3.1) can be rewritten in terms
of pseudo differential operators as follows:

$F(S_{t},t)=wS_{t}^{\gamma }\exp [-(T-t)(r+\psi (aD_{x}-i\gamma
))]I^{(w)}(\ln \frac{S_{t}}{K})$

\noindent where $I^{(w)}$ denotes the indicator function $%
I^{(w)}(x)=I_{[0,+\infty )}(wx)$. Here the notation $P(D_{x})$, with $%
D_{x}=-i\partial _{x}$ , denotes a pseudo differential operator whose symbol
is $P(\xi )$. Alternatively, $F(S_{t},t)=f(X_{t},t)$ can be viewed as a
solution to the pseudo differential equation: \ \ $[\partial _{t}-(r+\psi
(D_{x}))]f(X_{t},t)=0$ \ with the final condition $f(X_{T},T)=e^{\gamma
X_{T}}\mathbf{1}_{1}(waX_{T}\geq wK)$.\bigskip

\begin{lemma}
Let $F(S_{t},t)$ denote the current arbitrage-free price of the power option
with expiry $T$ payoff function $S_{T}^{\gamma }\mathbf{1}%
_{N}(w_{n}a_{n}X_{T}\geq w_{n}K_{n};n=1,...,N)$, $a_{n}\neq 0$, $w_{n}=\pm 1$%
, $X_{t}=\ln S_{t}.$ Then, for any $\omega _{n}>0$ such that $%
\tsum\limits_{n=1}^{N}w_{n}\omega _{n}a_{nk}+\gamma \in ]-\lambda
_{+},-\lambda _{-}[$,\ one can write\newline
\begin{equation}
F(S_{t},t)=\frac{e^{-r(T-t)}}{(2\pi i)^{N}}S_{t}^{\gamma
}\tprod\limits_{n=1}^{N}w_{n}\int\nolimits_{-\infty -iw_{N}\omega
_{N}}^{+\infty -iw_{N}\omega _{N}}...\int\nolimits_{-\infty -iw_{1}\omega
_{1}}^{+\infty -iw_{1}\omega _{1}}{}^{\text{ }}\frac{1}{\tprod%
\limits_{n=1,...N}\xi _{n}}.
\end{equation}%
\newline
$\ \ \ \ \ \ \ \ \ \ \ \ \ \ \ \ \ \ \ \ \ \ \ \ \ \ \ \ \ \ \ \ \ \ \ \ \ \
\ \ \ \ \ \ \ \ \ \ \ \ \ \ \ \ \ \ \exp [i\tsum\limits_{n=1}^{N}\xi
_{n}(a_{n}X_{t}-K_{n})-\Psi (t,\xi _{1},...\xi _{N})]d\xi _{1}..d\xi _{N}$%
\newline
where $\Psi (t,\xi _{1},...\xi _{N})=(T-t)\psi
(\tsum\limits_{n=1}^{N}a_{n}\xi _{n}-i\gamma )$
\end{lemma}

\noindent \textbf{Proof}. The result follows by writing the Fourier
transform of the payoff as a convolution of $N$ terms and by arguing as in
Lemma 1.\medskip

\noindent \textbf{Proof of Proposition 1. }Let us first prove the case $N=1$.%
\newline
Let \textbf{P}$=[(\gamma _{1},...\gamma _{M}),K,w,(a_{1},...a_{M})]$ with $%
w=\pm 1$. For $m=1,...,M$ \ let $K_{m}^{\ast
}=K-\tsum\nolimits_{k=1}^{m-1}a_{k}X_{k}$, $\gamma _{m}^{\ast
}=\tsum\nolimits_{k=1}^{m-1}\gamma _{k}X_{k}$ ($K_{1}^{\ast }=K$, $\gamma
_{m}^{\ast }=0$). Let $f_{m}(X_{t},t)$ solve $\partial _{t}f_{m}-(r+\psi
(D_{x}))f_{m}=0$ for $t\in \lbrack T_{m-1},T_{m}]$, with $%
f_{m}(T_{m},X_{m})=f_{m+1}(T_{m},X_{m})$ for $m<M$, and $%
f_{M}(T_{M},X_{M})=e^{\gamma _{M}X_{M}+\gamma _{M}^{\ast }}\mathbf{1}%
_{1}(wa_{M}X_{M}\geq wK_{M}^{\ast })$. In view of Lemma 2 and Remark 1 one
has:\newline
$f_{M}(X_{t},t)=\frac{we^{\gamma _{M}X_{t}+\gamma _{M}^{\ast }}}{2\pi i}%
\int\nolimits_{-\infty -iw\omega }^{+\infty -iw\omega }e^{i\xi \lbrack
a_{M}X_{t}-K_{M}^{\ast }]-(T_{M}-t)[r+\psi (a_{M}\xi -i\gamma _{M})]}\frac{1%
}{\xi }d\xi .$\newline
Then one can prove recursively that for any $m$:\newline
$f_{m}(X_{t},t)=\frac{we^{-r(T_{M}-t)}\exp (\gamma _{h}^{\ast
}+\tsum\limits_{k=m}^{M}\gamma _{k})}{2\pi i}\int\nolimits_{-\infty
-iw\omega }^{+\infty -iw\omega }\exp [i\xi
(\tsum\limits_{k=m}^{M}a_{k}X_{t}-K_{m}^{\ast })-\Psi _{m}(t,\xi )]\frac{1}{%
\xi }d\xi $\newline
where $\Psi _{m}(t,\xi )=\tsum\limits_{j=m}^{M}(T_{j}-T_{j-1})\psi (\xi
\tsum\limits_{k=j}^{M}a_{k}-i\tsum\limits_{k=j}^{M}\gamma _{k})\ \ $with $\
T_{m-1}=t,$ $\omega >0\ $and $\tsum\limits_{k=j}^{M}\left[ w\omega
a_{k}+\gamma _{k}\right] \in ]-\lambda _{+},-\lambda _{-}[$ $\ $ for $%
j=m,...,M$. Thus $h=1$ yields the result in the case $N=1$. Finally, the
general case is proved along the same lines, by employing Lemma 3 and
arguing recursively.\bigskip

In the sequel the pricing formulas for some exotics are obtained as an
application of Proposition 1. Note that our pricing formula are new in the L%
\'{e}vy environment. Moreover, our approach casts a new light also on the
Gaussian case that is obtained as a by-product.\bigskip

\section{Examples}

In this section we provide some examples to illustrate the flexibility of
the main formula. Most of the examples we give are new in the L\'{e}vy
framework. Note that we have chosen not to include in this section the
textbook examples and the most elementary examples concerning popular
options: some are given in [3]. A few lines in each example are devoted to
show how the well-known pricing expressions of the traditional Gaussian
setting can be obtained as special cases of ours quite straightforwardly.
Finally note that the references we give here are not exhaustive, because
the focus of this paper is on the analytical formulas, while the amount of
work on the purely computational issues is generally omitted.\medskip

\noindent 1) \textit{Forward-start options. }Forward-start options\textit{\ }%
are options that are only effective at some pre-specified time after they
have been bought or sold. They are the building-blocks of more complex
options. For example, a cliquet option is a series of consecutive
forward-start options, where each option becomes active when the former
expires. A valuation of forward-start options is given in [29] in the
classical Gaussian setting. Herewith the generalization to a L\'{e}vy
setting is presented.

Let $T$ \ be the expiration time and let $T_{1}<T$ \ be the time at which
the option starts. The strike price is set to be the underlying asset price
at $T_{1}$. Thus the final payoff is \ $\max \left\{
w(S_{T_{2}}-S_{T_{1}}),0\right\} $ with $w=\pm 1.$

As an application of Proposition 1 we give the current value of the
forward-start option $F(S_{t},t)$ in the above-mentioned L\`{e}vy setting,
under the assumptions $\lambda _{+}>0$ and $\lambda _{-}<-1.$

\noindent $F(S_{t},t)=$ $\frac{e^{-r(T_{2}-t)}}{2\pi i}S_{t}(\int%
\nolimits_{-\infty -iw\omega _{2}}^{+\infty -iw\omega _{2}}\frac{1}{\xi }%
\exp [-(T_{2}-T_{1})\psi (\xi -i)]d\xi $\newline
$-\int\nolimits_{-\infty -iw\omega _{1}}^{+\infty -iw\omega _{1}}\frac{1}{%
\xi }\exp [-(T_{2}-T_{1})\psi (\xi )]d\xi )$

\noindent where $\omega _{1}\in ]0,-\lambda _{-}[$ , $\omega _{2}\in
]0,-\lambda _{-}-1[$ if $w=1$ and $\omega _{1}\in ]0,\lambda _{+}[$ , $%
\omega _{2}\in ]0,\lambda _{+}+1[$ if $w=-1$.

\noindent Note that the greeks $\Delta $ (which is equal to $%
F(S_{t},t)/S_{t} $) and $\Gamma =0$ do not depend on $S_{t}.$ In the
Gaussian case, $\psi (\xi )=i(\frac{\sigma ^{2}}{2}-r)\xi +\frac{\sigma ^{2}%
}{2}\xi ^{2}$, the formula above becomes:

$\frac{1}{2\pi i}S_{t}(\int\nolimits_{-\infty -iw\omega _{2}}^{+\infty
-iw\omega _{2}}\frac{1}{\xi }\exp [i(r+\frac{\sigma ^{2}}{2}%
)(T_{2}-T_{1})\xi -\frac{(T_{2}-T_{1})\sigma ^{2}}{2}\xi ^{2}]d\xi $\newline
$-e^{-r(T_{2}-T_{1})}\int\nolimits_{-\infty -iw\omega _{1}}^{+\infty
-iw\omega _{1}}\frac{1}{\xi }\exp [i(r-\frac{\sigma ^{2}}{2}%
)(T_{2}-T_{1})\xi -\frac{(T_{2}-T_{1})\sigma ^{2}}{2}\xi ^{2}]d\xi )$

\noindent which is $wS_{t}[N(w(\frac{r}{\sigma }+\frac{\sigma }{2})\sqrt{%
T_{2}-T_{1}})-e^{-r(T_{2}-T_{1})}N(w(\frac{r}{\sigma }-\frac{\sigma }{2})%
\sqrt{T_{2}-T_{1}})$, i.e. the formula found in [29].\medskip

\textit{2)} \textit{Asian options}. Asian options under L\`{e}vy processes
have been priced in a number of papers (see [5], [18], [21]). In this
subsection we show how a pricing formula for geometric Asian options is
easily obtained from our general result. At first we consider discrete Asian
options - whose payoff depends on a discrete average of the asset price at $%
N $ monitoring times, $T_{1}<...<T_{M}$ . The continuous average case is
obtained as a limit. Consider a forward-start fixed strike Asian option with
strike price $K.$ Note that a pricing expression for floating Asian options
is easily obtained in view of the symmetry relationship proved in [14]. Let $%
T=T_{M}$ be the maturity date and let $T^{\prime }=T_{1}$ be the time at
which the averaging starts. The payoff is $\max (w[\Sigma _{M}-K],0)$, where 
$\Sigma _{M}=(\tprod\limits_{j=1,...,M}S_{T_{j}})^{\frac{1}{M}}$ and $w$ is
the binary indicator. In terms of $X_{t}=\ln (S_{t})$ the payoff is:\newline
$w\tprod\limits_{j=1,...,M}e^{\frac{1}{M}X_{T_{j}}}\mathbf{1}_{1}(wA$\textbf{%
X}$\geq w\ln (K))-wK\mathbf{1}_{1}(wA$\textbf{X}$\geq w\ln (K))$, with $A=[%
\frac{1}{M},...,\frac{1}{M}]$.\newline
Thus Proposition 1 applies with $N=1$ and yields the following valuation
formula, after some algebraic manipulation:

$F(S_{t},t)=-\frac{Ke^{-r(T_{M}-t)}}{2\pi }\int\nolimits_{-\infty -iw\omega
}^{+\infty -iw\omega }\frac{1}{\xi (\xi +i)}\exp [i\xi \ln \frac{S_{t}}{K}%
-\tsum\limits_{j=1}^{M}(T_{j}-T_{j-1})\psi (\xi \frac{M-j}{M})]d\xi $

\noindent with $\ T_{0}=t,$ $\omega \in ]1,-\lambda _{-}[$ \ ($]0,\lambda
_{+}[$) if $w=1$ ($w=-1$). Note that the analogous expression obtained in
[18], (10), is derived throughout a different argument, that is considering
the distribution of $\ln (\Sigma _{M}).$

The pricing formula for the continuous-time monitoring case, where the
geometric average is $\exp [\frac{1}{T-T^{\prime }}\tint\nolimits_{T^{\prime
}}^{T}\ln (S_{t})dt]$, follows from the discrete pricing formula just
letting $M\rightarrow \infty .$ Note that the limit can be computed under
the integral sign in view of the nice behavior of $\psi .$ In particular,
for the continuous case, one has:\medskip

$F(S_{t},t)=-\frac{Ke^{-r(T-t)}}{2\pi }\int\nolimits_{-\infty -iw\omega
}^{+\infty -iw\omega }\frac{1}{\xi (\xi +i)}\exp [i\xi \ln \frac{S_{t}}{K}%
-\tint\nolimits_{0}^{1}\psi (\xi (1-y))dy]d\xi .\medskip $

Let us now see that our formula collapses to the know valuation formula for
discretely monitored Asian options in the Gaussian case (see [29]). Let $h$
denote the averaging frequency, that is, $T_{j}=T-(M-j)h$, $j=1,...,M$.
Then, in the Gaussian case, $\tsum\limits_{j=1}^{M}(T_{j}-T_{j-1})\psi (\xi 
\frac{M-j}{M})=-ih(r-\frac{\sigma ^{2}}{2})\xi \frac{M-1}{2}+h\frac{\sigma
^{2}}{2}\xi ^{2}\frac{(M-1)(2M-1)}{6M}$, because $\tsum%
\limits_{j=1}^{M}(M-j)=\frac{M(M-1)}{2}$ and $\tsum%
\limits_{j=1}^{M}(M-j)^{2}=\frac{M(M-1)(2M-1)}{6}.$ Thus\newline
$F(S_{t},t)=\frac{Ke^{-r(T-t)}}{2\pi i}\int\nolimits_{-\infty -iw\omega
}^{+\infty -iw\omega }\exp [i\xi (\ln \frac{S_{t}}{K}+\frac{1}{2}(r-\frac{%
\sigma ^{2}}{2})(T-T^{\prime }))-(T-T^{\prime })\frac{\sigma ^{2}}{2}\xi ^{2}%
\frac{2M-1}{6M}]$\newline
$\ \ \ \ \ \ \ \ \ \ \ \ \ \ \ \ \ \ \ \ \ \ \ \ \ \ \ .(\frac{1}{\xi +i}-%
\frac{1}{\xi })d\xi $

\noindent for any $\omega >0.$ Splitting the integral into two integrals and
changing variables, one gets:\newline
$F(S_{t},t)=\frac{S_{t}e^{-\beta }}{2\pi i}\int\nolimits_{-\infty -iw\omega
}^{+\infty -iw\omega }\exp [i\eta D^{+}-\frac{\sigma ^{2}}{2}\eta ^{2}]\frac{%
1}{\eta }d\eta -\frac{Ke^{-r(T-t)}}{2\pi i}\int\nolimits_{-\infty -iw\omega
}^{+\infty -iw\omega }\exp [i\eta D^{-}-\frac{\sigma ^{2}}{2}\eta ^{2}]\frac{%
1}{\eta }d\eta $

\noindent where $D^{-}=[\ln \frac{S_{t}}{K}+\frac{1}{2}(r-\frac{\sigma ^{2}}{%
2})(T-T^{\prime })]/(\sigma \sqrt{T-T^{\prime }}\sqrt{\frac{2M-1}{6M}}),$%
\newline
$D^{+}=D^{-}+\sigma \sqrt{T-T^{\prime }}\sqrt{\frac{2M-1}{6M}}$ and $\beta
=r(T-t)+(r+\sigma ^{2}(\frac{1}{2}-\frac{2M-1}{6M}))\frac{T-T^{\prime }}{2}$%
, which yields the known formula (see [29]) by application of Lemma 1.$%
\medskip $

Finally we point out that Proposition 1 straightforwardly applies to the
more general flexible geometric Asian options, where the flexible geometric
average is \ $\tprod\limits_{j=1,...,M}S_{T_{j}}^{\theta _{j}}$ with $\theta
_{j}=\theta (j)/\tsum\nolimits_{j=1}^{M}\theta (j)$ and $\theta $ any
non-negative function (see [29] for the Gaussian case). The following
expression is obtained:\newline
$F(S_{t},t)=-\frac{Ke^{-r(T_{M}-t)}}{2\pi }\int\nolimits_{-\infty -iw\omega
}^{+\infty -iw\omega }\frac{1}{\xi (\xi +i)}\exp [i\xi \ln \frac{S_{t}}{K}%
-\tsum\limits_{j=1}^{M}(T_{j}-T_{j-1})\psi (\xi \tsum\limits_{k=j}^{M}\theta
_{k})]d\xi $

\noindent with $\ T_{0}=t,$ $\omega \in ]1,-\lambda _{-}[$ \ ($]0,\lambda
_{+}[$) if $w=1$ ($w=-1$).\bigskip

\noindent 3) \textit{Discrete lookback options}. Lookback options are
path-dependent options whose payoff depends on the extremal price of the
underlying asset over the life of the option. We focus on the realistic case
of finite sampling lookback, that is, the asset price is monitored at
particular dates. Moreover, we confine ourselves to fixed strikes options,
that is, the payoff is the maximum difference between the optimal price and
the strike price $K$ which is determined at inception. According to [12]
"perhaps, in exponential L\`{e}vy model closed-form formulas are not, in
general, available for pricing these options". In the traditional Gaussian
setting discrete lookback options have been priced in [20] exactly, and in
[8] by an adjustment of the continuous case.

Let $T_{1}<T_{2}<...T_{M}$ be the monitoring times and assume that the
lookback period $[T_{1},T_{N}]$ starts in the future, i.e. the current time $%
t<T_{1}$, and ends at expiration, i.e. $T_{M}=T$, the expiration date.
However our method applies also to backward starting options. The payoff is $%
\max (S_{T_{1}},...,S_{T_{M}},K)-K$ for the call and $K-\min
(S_{T_{1}},...,S_{T_{M}},K)$ for the put, that is, $\max
(wS_{T_{1}},...,wS_{T_{M}},wK)-wK$ for any option, where $w=\pm 1$.
Following [20] one can write the payoff in our notation as follows:

\noindent $\tsum\limits_{p=1}^{M}we^{X_{p}}[$\textbf{1}$_{M-1}(wA^{(p)}%
\mathbf{X}>0)-$\textbf{1}$_{M}(w\widetilde{A}^{(p)}\mathbf{X}%
>wB^{(p)})]-wK(1-$\textbf{1}$_{M}(w\mathbf{X}<w\widehat{K}))$\newline
where

\noindent $A^{(p)}$ is an $(M-1)\times M$ matrix whose entries are $%
A_{ij}^{(p)}=\left\{ 
\begin{array}{lll}
1 & if & j=p \\ 
-1 & if & i=j\neq p \\ 
0 &  & otherwise%
\end{array}%
\right. $;

\noindent $\widetilde{A}^{(p)}$ is an $M\times M$ matrix whose entries are $%
\widetilde{A}_{ij}^{(p)}=\left\{ 
\begin{array}{lll}
1 & if & j=p,i\neq j \\ 
-1 & if & i=j \\ 
0 &  & otherwise%
\end{array}%
\right. $;

\noindent $B^{(p)}$ is an $M-$dimension vector whose entries are $0$ with
the exception of the $p^{th}$ that is $-\ln K$;

\noindent $\widehat{K}$ is an $M-$dimension vector whose entries are all $%
-\ln K$.

\noindent Thus Proposition 1 yields the following expression for the current
value $F(S_{t},t)$ of the lookback option:\medskip

\noindent (4.1) $\ \ \ \frac{w^{M}S_{t}e^{-r(T_{M}-t)}}{(2\pi i)^{M-1}}%
\tsum\limits_{p=1}^{M}\int\nolimits_{-\infty -iw\omega _{1}}^{+\infty
-iw\omega _{1}}...\int\nolimits_{-\infty -iw\omega _{M}}^{+\infty -iw\omega
_{M}}\frac{1}{\tprod\limits_{\substack{ n=1,...N  \\ n\neq p}}\xi _{n}}\exp
[-\tsum\limits_{j=1}^{p}(T_{j}-T_{j-1})$

$\ \ \ \ \ \ \ \ \ \ \ \psi (\tsum\limits_{n<j}\xi
_{n}-i)]-\tsum\limits_{j=p+1}^{M}(T_{j}-T_{j-1})\psi (-\tsum\limits_{n\geq
j}\xi _{n})]d\xi _{1}...d\xi _{p-1}d\xi _{p+1}...d\xi _{M}$

$-\frac{w^{M}S_{t}e^{-r(T_{M}-t)}}{(2\pi i)^{M-1}}\tsum\limits_{p=1}^{M}\int%
\nolimits_{-\infty -iw\omega _{1}}^{+\infty -iw\omega
_{1}}...\int\nolimits_{-\infty -iw\omega _{M}}^{+\infty -iw\omega _{M}}\frac{%
1}{\tprod\limits_{\substack{ n=1,...N  \\ n\neq p}}\xi _{n}}\exp
[-\tsum\limits_{j=1}^{p}(T_{j}-T_{j-1})$

$\ \ \ \ \ \ \ \ \ \psi (\tsum\limits_{n<j}\xi
_{n}-i)-\tsum\limits_{j=p+1}^{M}(T_{j}-T_{j-1})\psi (-\tsum\limits_{n\geq
j}\xi _{n})]d\xi _{1}...d\xi _{p-1}d\xi _{p+1}...d\xi _{M}\bigskip $

$-\frac{w^{M+1}S_{t}e^{-r(T_{M}-t)}}{(2\pi i)^{M}}\tsum\limits_{p=1}^{M}\int%
\nolimits_{-\infty -iw\omega _{1}}^{+\infty -iw\omega
_{1}}...\int\nolimits_{-\infty -iw\omega _{M}}^{+\infty -iw\omega _{M}}\frac{%
1}{\tprod\limits_{n=1,...N}\xi _{n}}\exp [-i\xi _{p}\ln \frac{S_{t}}{K}$

$-\tsum\limits_{j=1}^{p}(T_{j}-T_{j-1})\psi (\tsum\limits_{n<j}\xi _{n}-\xi
_{p}-i)-\tsum\limits_{j=p+1}^{M}(T_{j}-T_{j-1})\psi (-\tsum\limits_{n\geq
j}\xi _{n})]d\xi _{1}...d\xi _{M}\bigskip $

$+\frac{w^{M+1}Ke^{-r(T_{M}-t)}}{(2\pi i)^{M}}\int\nolimits_{-\infty
-iw\omega _{1}}^{+\infty -iw\omega _{1}}...\int\nolimits_{-\infty -iw\omega
_{M}}^{+\infty -iw\omega _{M}}\frac{1}{\tprod\limits_{n=1,...N}\xi _{n}}\exp
[-i\tsum\limits_{n=1}^{M}\xi _{n}\ln \frac{S_{t}}{K}-$

$\ \ \ \ \ \tsum\limits_{j=1}^{p}(T_{j}-T_{j-1})\psi
(-\tsum\limits_{n=j}^{M}\xi _{n})]d\xi _{1}...d\xi _{M}$ $%
-wKe^{-r(T_{M}-t)}\bigskip $

\noindent with some positive $\omega _{n}$ such that such that $%
\tsum\limits_{n=1}^{M}\omega _{n}<\min (\lambda _{+},-\lambda _{-}-1)$.

Let us see how our general formula collapses to (21) of [20]. For any $%
p=1,...,M$ denote:\newline
$q_{p}^{\pm }=[\ln \frac{S_{t}}{K}+(r\pm \frac{\sigma ^{2}}{2}%
)(T_{p}-t)]/(\sigma \sqrt{T_{p}-t})$, $g_{pn}^{\pm }=(\frac{r}{\sigma }\pm 
\frac{\sigma }{2})\sqrt{T_{p}-T_{n}}$ for $n=1,...,p-1$, $h_{pn}=(\frac{%
\sigma }{2}-\frac{r}{\sigma })\sqrt{T_{n}-T_{p}}$ for $n=p+1,...,M$. Split
any integral in the first sum into two integrals, the former in $d\xi
_{1}...d\xi _{p-1}$ and the latter in $d\xi _{p+1}...d\xi _{M}$, and then
change to variables $\xi _{n}^{\prime }\sigma \sqrt{T_{p}-T_{n}}=\xi _{n}$ \
($\xi _{n}^{\prime }\sigma \sqrt{T_{n}-T_{p}}=\xi _{n}$, respectively). Then
the integrals in the first sum of (4.1) becomes the product of:\medskip 
\newline
$\frac{w^{p}S_{t}e^{-r(T_{M}-T_{p})}}{(2\pi i)^{p-1}}\int\nolimits_{-\infty
-iw\omega _{1}}^{+\infty -iw\omega _{1}}...\int\nolimits_{-\infty -iw\omega
_{p-1}}^{+\infty -iw\omega _{p-1}}\frac{1}{\tprod\limits_{n=1,...p}\xi _{n}}%
\exp [\tsum\limits_{n=1}^{p-1}(i\xi _{n}g_{pn}^{+}-\frac{1}{2}\xi _{n}^{2}-$

$\ \ \ \ \ \ \ \ \ \ \ \ \ \ \ \ \ \ \ \ \ \ \ \ \ \ \ \ \ \ \ \ \ \ \
\tsum\limits_{m<n}\xi _{n}\xi _{m}\sqrt{\frac{T_{p}-T_{n}}{T_{p}-T_{m}}}%
)]d\xi _{1}...d\xi _{p-1}$\newline
and\newline
$\frac{w^{M-p}}{(2\pi i)^{M-p}}\int\nolimits_{-\infty -iw\omega
_{p+1}}^{+\infty -iw\omega _{p+1}}...\int\nolimits_{-\infty -iw\omega
_{M}}^{+\infty -iw\omega _{M}}\frac{1}{\tprod\limits_{n=p+1,...M}\xi _{n}}%
\exp [\tsum\limits_{n=p+1}^{M}(i\xi _{n}h_{pn}-\frac{1}{2}\xi _{n}^{2}-$

$\ \ \ \ \ \ \ \ \ \ \ \ \ \ \ \ \ \ \ \ \ \ \ \ \ \ \ \ \ \ \ \ \ \ \
\tsum\limits_{m>n}\xi _{n}\xi _{m}\sqrt{\frac{T_{n}-T_{p}}{T_{m}-T_{p}}}%
)]d\xi _{p+1}...d\xi _{M}.$\newline
Employing Lemma 1 each term in the first sum is transformed into:\newline
$wS_{t}e^{-r(T_{M}-T_{p})}N_{p-1}(wg_{p1}^{+},...,wg_{p,p-1}^{+};\Delta
_{p})N_{M-p}(wh_{p,p+1},...,wh_{pM};\Theta _{p})$\newline
where the correlation matrix $\Delta _{p}$ has typical element $\sqrt{\frac{%
T_{p}-T_{n}}{T_{p}-T_{m}}}$ , $m<n$, and the correlation matrix $\Theta _{p}$
has typical element $\sqrt{\frac{T_{n}-T_{p}}{T_{m}-T_{p}}}$, $m>n$. A
similar treatment on the terms in the second sum turns each term into:%
\newline
$wS_{t}e^{-r(T_{M}-T_{p})}N_{p}(wg_{p1}^{+},...,wg_{p,p-1}^{+},-wq_{p}^{+};%
\Delta _{p}^{\ast })N_{M-p}(wh_{p,p+1},...,wh_{pM};\Theta _{p})$\newline
where the correlation matrix $\Delta _{p}^{\ast }$ has typical element $\rho
_{mn}^{\ast }=\sqrt{\frac{T_{p}-T_{n}}{T_{p}-T_{m}}}$ for $n>m$, $%
m=1,...,p-1 $, $\rho _{pn}^{\ast }=\sqrt{\frac{T_{p}-T_{n}}{T_{p}-t}}$ for $%
n<p$, and $\rho _{nn}^{\ast }=1$ for any $n$. Finally, the integrals in the
third sum are turned into \ $%
wKe^{-r(T_{M}-t)}N_{M}(-wq_{1}^{-},...,-wq_{M}^{+};\widehat{\Delta })$ \
where the correlation matrix has typical element $\widehat{\rho }_{hk}=\frac{%
\min (T_{h}-t,T_{k}-t)}{\sqrt{T_{h}-t}\sqrt{T_{k}-t}}$.\bigskip

4) \textit{Chooser options}

\noindent A chooser option gives its holder the right to decide at a
prespecified time (choice date = $T_{1}$) before the maturity $T$ \ whether
he/she would like the option to be a call or a put option. As a
straightforward application of Proposition 1 (with $A=I$) we give a
valuation formula for simple chooser options, i.e. the call and the put have
the same strike price $K$ and maturity date $T.$ Note that the decision
whether the option is a call or a put depends on the value of:\newline
$%
Max(C(S_{T_{1}};K,T),P(S_{T_{1}};K,T))=C(S_{T_{1}};K,T)+Max(Ke^{-r(T-T_{1})}-S_{T_{1}},0). 
$

\noindent In other words the choice is:\newline
Call \ $\Longleftrightarrow Ke^{-r(T-T_{1})}<S_{T_{1}}$ ; \ Put \ $%
\Longleftrightarrow Ke^{-r(T-T_{1})}>S_{T_{1}}.$

\textbf{\noindent }The\textbf{\ }payoff can be expressed as:\newline
$(S_{T}-K)$\textbf{1}$_{2}(S_{T}>K,S_{T_{1}}>Ke^{-r(T-T_{1})})+(K-S_{T_{1}})$%
\textbf{1}$_{2}(S_{T}<K,S_{T_{1}}<Ke^{-r(T-T_{1})})$

\noindent Then, in view of Proposition 1, with $N=1$ and $M=2$, the price
for the simple chooser option at time $t<T_{1}$ can be written in the form:$%
\medskip $

$F(S_{t},t)=A_{1}-K\ast A_{2}+K\ast A_{3}-A_{4}\medskip $

\noindent where the following choice are made in Proposition 1 for each term:

\noindent for $\ A_{1}:$ \ \textbf{K}$=(Ke^{-r(T-T_{1})},K)$ \ \ \ \ \ \ $%
\gamma =(0,1)$ \ \ \ \ \ \ \ \ \ $(w_{1},w_{2})=(1,1)$\newline
for $\ A_{2}:$ \ \textbf{K}$=(Ke^{-r(T-T_{1})},K)$ \ \ \ \ \ \ $\gamma
=(0,0) $ \ \ \ \ \ \ \ \ \ $(w_{1},w_{2})=(1,1)$\newline
for $\ A_{3}:$ \ \textbf{K}$=(Ke^{-r(T-T_{1})},K)$ \ \ \ \ \ \ $\gamma
=(0,0) $ \ \ \ \ \ \ \ \ \ $(w_{1},w_{2})=(-1,-1)$\newline
for $\ A_{4}:$ \ \textbf{K}$=(Ke^{-r(T-T_{1})},K)$ \ \ \ \ \ \ $\gamma
=(0,1) $ \ \ \ \ \ \ \ \ \ $(w_{1},w_{2})=(-1,-1)$.

Then\newline
$A_{1}=\frac{S_{t}e^{-r(T-t)}}{(2\pi i)^{2}}\int\nolimits_{-\infty -i\omega
_{2}}^{+\infty -i\omega _{2}}$ $\int\nolimits_{-\infty -i\omega
_{1}}^{+\infty -i\omega _{1}}e^{i\xi _{1}\ln (\frac{S_{t}}{K}%
+r(T-T_{1})+i\xi _{2}\ln \frac{S_{t}}{K}-\Psi (t,\xi _{1},\xi _{2}-i)}\frac{1%
}{\xi _{1}\xi _{2}}d\xi _{1}d\xi _{2}$\newline
In view of the residue theorem $A_{1}$ becomes:\newline
$=A_{4}+\frac{S_{t}e^{-r(T-t)}}{2\pi i}\int\nolimits_{-\infty -i\omega
_{2}}^{+\infty -i\omega _{2}}$ $e^{i\xi _{2}\ln \frac{S_{t}}{K}-(T-t)\psi
(\xi _{2}-i)}\frac{1}{\xi _{2}}d\xi _{2}+$\newline
$\ \ \ \ \ \ +\frac{S_{t}e^{-r(T-t)}}{2\pi i}\int\nolimits_{-\infty +i\omega
_{1}}^{+\infty +i\omega _{1}}e^{i\xi _{1}\ln (\frac{S_{t}}{K}%
+r(T-T_{1})-(T_{1}-t)\psi (\xi _{1}-i)}\frac{1}{\xi _{1}}d\xi _{1}$

\noindent On the other hand

\noindent $A_{2}=\frac{e^{-r(T-t)}}{(2\pi i)^{2}}\int\nolimits_{-\infty
+i\omega _{2}}^{+\infty +i\omega _{2}}$ $\int\nolimits_{-\infty +i\omega
_{1}}^{+\infty +i\omega _{1}}e^{i\xi _{1}\ln (\frac{S_{t}}{K}%
+r(T-T_{1})+i\xi _{2}\ln \frac{S_{t}}{K}-\Psi (t,\xi _{1},\xi _{2})}\frac{1}{%
\xi _{1}\xi _{2}}d\xi _{1}d\xi _{2}$\newline
which under the residue theorem is transformed into:\newline
$=A_{3}+\frac{e^{-r(T-t)}}{2\pi i}\int\nolimits_{-\infty +i\omega
_{2}}^{+\infty +i\omega _{2}}$ $e^{i\xi _{2}\ln \frac{S_{t}}{K}-(T-t)\psi
(\xi _{2})}\frac{1}{\xi _{2}}d\xi _{2}+$\newline
$\ \ \ \ \ +\frac{e^{-r(T-t)}}{2\pi i}\int\nolimits_{-\infty -i\omega
_{1}}^{+\infty -i\omega _{1}}e^{i\xi _{1}\ln (\frac{S_{t}}{K}%
+r(T-T_{1})-(T_{1}-t)\psi (\xi _{1})}\frac{1}{\xi _{1}}d\xi _{1}\bigskip $

\noindent Thus the formula is simplified because the double integrals cancel
out and the final expression is:

\noindent $F(S_{t},t)=\frac{S_{t}e^{-r(T-t)}}{2\pi i}[\int\nolimits_{-\infty
-i\omega _{2}}^{+\infty -i\omega _{2}}$ $e^{i\xi _{2}\ln \frac{S_{t}}{K}%
-(T-t)\psi (\xi _{2}-i)}\frac{1}{\xi _{2}}d\xi _{2}+$\newline
\newline
$\ \ \ \ \ \ \ \ \ \ \ \ \ \ \ \ \ \ \ \ \ \int\nolimits_{-\infty +i\omega
_{1}}^{+\infty +i\omega _{1}}e^{i\xi _{1}\ln (\frac{S_{t}}{K}%
+r(T-T_{1})-(T_{1}-t)\psi (\xi _{1}-i)}\frac{1}{\xi _{1}}d\xi _{1}]-$\newline
\newline
\ \ \ \ \ \ \ \ \ \ \ \ \ \ \ \ \ \ \ \ \ \ $\frac{Ke^{-r(T-t)}}{2\pi i}%
[\int\nolimits_{-\infty +i\omega _{2}}^{+\infty +i\omega _{2}}$ $e^{i\xi
_{2}\ln \frac{S_{t}}{K}-(T-t)\psi (\xi _{2})}\frac{1}{\xi _{2}}d\xi _{2}+$%
\newline
$\ \ \ \ \ \ \ \ \ \ \ \ \ \ \ \ \ \ \ \ \ \ \ \int\nolimits_{-\infty
-i\omega _{1}}^{+\infty -i\omega _{1}}e^{i\xi _{1}\ln (\frac{S_{t}}{K}%
+r(T-T_{1})-(T_{1}-t)\psi (\xi _{1})}\frac{1}{\xi _{1}}d\xi _{1}\bigskip ].$

\noindent In the Gaussian case the formula becomes\medskip \newline
$%
F(S_{t},t)=S[N(d_{T}^{+})-N(-d_{T,T_{1}}^{+})]-Ke^{-r(T-t)}[N(d_{T}^{-})-N(-d_{T,T_{1}}^{-})] 
$\newline
with $d_{T}^{\pm }=(\ln (S_{t}/K)+(r\pm \frac{\sigma ^{2}}{2})(T-t))/(\sigma 
\sqrt{T-t})$\newline
\ \ \ \ \ \ $d_{T,T_{1}}^{+}=d_{T_{1}}^{\pm }+r(T-T_{1})/(\sigma \sqrt{%
T_{1}-t})\medskip $

\noindent which is the price for a chooser option obtained by Rubinstein in
1991 (See also [29]).\bigskip

\textit{5) Multicompound options}. A notable example of application of our
Proposition is the valuation of multicompound options (of order $N$), that
is options whose underlying asset is a multicompound option of order $N-1$.
A closed form expression for this kind of options was developed by [19] in
the usual Gaussian setting. An extension to a more general L\'{e}vy
environment has been provided in [3] for the compound call options of order
2. Here we give a valuation expression which holds for any order. Note that
multicompound options have been introduced to face the problem of pricing
defaultable coupon bonds ([19], [4]) ; however they have a wide range of
applications to all opportunities having a sequential nature (see [1], [2]
for applications in real option analysis).

Let $t<T_{1}<T_{2}<...T_{N}$ and let $%
F_{N}(S_{t},t;T_{1},K_{1},w_{1};...;T_{N},K_{N},w_{N})$ denote the current
value of a European $N$-fold compound option expiring at time $T_{1}$, with
strike price $K_{1}$ and with as underlying asset a European $(N-1)$-fold
compound option expiring at time $T_{2}$, with strike price $K_{2}$,
....until the final underlying asset, a European option on a stock, with
exercise date and price given by $T_{N}$ and $K_{N}$. As usually $w_{j}=\pm
1 $ represents the call/put attribute of each option. Assume that the
stochastic process followed by the underlying stock is a RLPE. Suppose that $%
S_{j}^{\ast }$ is the solution to $F_{N-j}(S_{j}^{\ast
},T_{j};T_{j+1},K_{j+1},w_{j+1};...;T_{N},K_{N},w_{N})=K_{j}$ for $%
j=1,...,N-1$. We will comment on existence and uniqueness of $S_{j}^{\ast }$
later on. Then the payoff of the multicompound option can be written as:

$\tprod\limits_{n=1}^{N}w_{n}S_{T_{N}}$\textbf{1}$_{N}(w_{N}S_{T_{N}}\geq
w_{N}S_{N}^{\ast },...,w_{1}S_{T_{1}}\geq w_{1}S_{1}^{\ast
})-\tsum\limits_{j=1}^{N}\tprod\limits_{n=1}^{j}w_{n}K_{j}$\textbf{1}$%
_{j}(w_{j}S_{T_{j}}\geq w_{j}S_{j}^{\ast },...,w_{1}S_{T_{1}}\geq
w_{1}S_{1}^{\ast })$

\noindent where $S_{N}^{\ast }=K_{N}$ therefore our method applies. The
current value of each term is obtained by straightforward application of
Proposition 1 and the resulting expression reads:

$\overset{}{%
\begin{array}{c}
\frac{S_{t}e^{-r(T_{M}-t)}}{(2\pi i)^{N}}\int\nolimits_{-\infty
-iw_{1}\omega _{1}}^{+\infty -iw_{1}\omega _{1}}...\int\nolimits_{-\infty
-iw_{N}\omega _{N}}^{+\infty -iw_{N}\omega _{N}}{}^{\text{ }}\frac{1}{%
\tprod\limits_{n=1,...N}\xi _{n}}\exp [i\tsum\limits_{n=1}^{N}\xi _{n}\ln 
\frac{S_{t}}{S_{n}^{\ast }}] \\ 
.\exp [-\Psi _{N}(t,\xi _{1},..,\xi _{N}-i)]d\xi _{1}...d\xi _{N}- \\ 
\newline
\tsum\limits_{j=1}^{N}\frac{K_{j}e^{-r(T_{j}-t)}}{(2\pi i)^{j}}\
\int\nolimits_{-\infty -iw_{1}\omega _{1}}^{+\infty -iw_{1}\omega
_{1}}...\int\nolimits_{-\infty -iw_{j}\omega _{j}}^{+\infty -iw_{j}\omega
_{j}}{}^{\text{ }}\frac{1}{\tprod\limits_{n=1,...j}\xi _{n}}\exp
[i\tsum\limits_{n=1}^{j}\xi _{n}\ln \frac{S_{t}}{S_{n}^{\ast }}] \\ 
.\exp [-\Psi _{j}(t,\xi _{1},..,\xi _{j})]d\xi _{1}...d\xi _{j}%
\end{array}%
}$

\noindent for some positive $\omega _{n}$ such that $\tsum%
\limits_{n=1}^{j}w_{n}\omega _{n}\in ]-\lambda _{+},-\lambda _{-}-1[$ $%
\forall j=1,...,N$, and where $\Psi _{j}(t,\xi _{1},..,\xi
_{j})=\tsum\limits_{k=1}^{j}(T_{k}-T_{k-1})\psi (\tsum\limits_{n=1}^{j}\xi
_{n})$.

Note that, differentiating under the integral sign, one can prove that $%
\partial _{h_{j}}F_{N}=0$ for $h_{j}=\frac{S_{t}}{S_{j}^{\ast }}$, $%
j=1,...,N $. Thus $\partial _{S_{t}}F_{N}$ is just the first integral of $%
F_{N}$ divided by $S_{t}$. Therefore uniqueness of $S_{j}^{\ast }$ is
guaranteed for any $j$. Existence holds for the multicompound call options,
while in the general case it holds only for suitable strike prices.

In the Gaussian case, if we change variables $\xi _{j}\sigma \sqrt{T_{j}-t}%
=\eta _{j}$ , let $\omega _{j}$ be any positive number and denote $(\ln 
\frac{S_{t}}{S_{j}^{\ast }}+(r\pm \frac{\sigma ^{2}}{2})(T_{j}-t))/(\sigma 
\sqrt{T_{j}-t})$ \ by $\ d_{j}^{\pm }$, and $\sqrt{\frac{T_{j}-t}{T_{k}-t}}$
by $\rho _{jk}$ if $j<k$, then our formula collapses into:\newline
$S_{t}\tprod\limits_{n=1}^{N}w_{n}N_{N}(w_{N}d_{N}^{+},...,w_{1}d_{1}^{+};%
\Xi
_{N})-\tsum\limits_{j=1}^{N}\tprod%
\limits_{n=1}^{j}w_{n}K_{j}N_{j}(w_{j}d_{j}^{-},...,w_{1}d_{1}^{+};\Xi _{j})$%
\newline
where $\Xi _{j}$ is a $j\times j-$ matrix with typical elements $\rho
_{jk}w_{k}w_{j}$ when $j<k$.\bigskip

\noindent \textit{Remark. (Put-call parity for compound options). }Note
that, starting from our formula, one can also verify the put-call parity
relationship for compound options as a nice exercise. Indeed\newline
$F_{2}(S_{t},t;1,K_{1},T_{1};w_{2},K_{2},T_{2})=$\newline
$=\frac{S_{t}e^{-r(T_{2}-t)}}{(2\pi i)^{2}}\int\nolimits_{-\infty
-iw_{2}\omega _{2}}^{+\infty -iw_{2}\omega _{2}}$ $\int\nolimits_{-\infty
-i\omega _{1}}^{+\infty -i\omega _{1}}e^{i\xi _{1}\ln \frac{S_{t}}{S\ast }%
+i\xi _{2}\ln \frac{S_{t}}{K_{2}}-\Psi (t,\xi _{1},\xi _{2}-i)}\frac{1}{\xi
_{1}\xi _{2}}d\xi _{1}d\xi _{2}-$\newline
$\frac{K_{2}e^{-r(T_{2}-t)}}{(2\pi i)^{2}}\int\nolimits_{-\infty
-iw_{2}\omega _{2}}^{+\infty -iw_{2}\omega _{2}}$ $\int\nolimits_{-\infty
-i\omega _{1}}^{+\infty -i\omega _{1}}e^{i\xi _{1}\ln \frac{S_{t}}{S\ast }%
+i\xi _{2}\ln \frac{S_{t}}{K_{2}}-\Psi (t,\xi _{1},\xi _{2})}\frac{1}{\xi
_{1}\xi _{2}}d\xi _{1}d\xi _{2}-$\newline
$\frac{K_{1}e^{-r(T_{1}-t)}}{2\pi }\int\nolimits_{-\infty -i\omega
_{1}}^{+\infty -i\omega _{1}}e^{i\xi _{1}\ln \frac{S_{t}}{S\ast }%
-(T_{1}-t)\psi (\xi _{1})}\frac{1}{i\xi _{1}}d\xi _{1}.\bigskip $

\noindent Let us shift the line of integration $\func{Im}\xi _{1}=\omega
_{1} $ up. Since we cross the pole at $\xi _{1}=0$, the residue theorem
gives:

\noindent $F_{2}(S_{t},t;1,K_{1},T_{1};w_{2},K_{2},T_{2})=$\newline
$\frac{S_{t}e^{-r(T_{2}-t)}}{(2\pi i)^{2}}\int\nolimits_{-\infty
-iw_{2}\omega _{2}}^{+\infty -iw_{2}\omega _{2}}$ $\int\nolimits_{-\infty
+i\omega _{1}}^{+\infty +i\omega _{1}}e^{i\xi _{1}\ln \frac{S_{t}}{S\ast }%
+i\xi _{2}\ln \frac{S_{t}}{K_{2}}-\Psi (t,\xi _{1},\xi _{2}-i)}\frac{1}{\xi
_{1}\xi _{2}}d\xi _{1}d\xi _{2}-$\newline
$\frac{K_{2}e^{-r(T_{2}-t)}}{(2\pi i)^{2}}\int\nolimits_{-\infty
-iw_{2}\omega _{2}}^{+\infty -iw_{2}\omega _{2}}$ $\int\nolimits_{-\infty
+i\omega _{1}}^{+\infty +i\omega _{1}}e^{i\xi _{1}\ln \frac{S_{t}}{S\ast }%
+i\xi _{2}\ln \frac{S_{t}}{K_{2}}-\Psi (t,\xi _{1},\xi _{2})}\frac{1}{\xi
_{1}\xi _{2}}d\xi _{1}d\xi _{2}-$ \newline
$\frac{K_{1}e^{-r(T_{1}-t)}}{2\pi }\int\nolimits_{-\infty +i\omega
_{1}}^{+\infty +i\omega _{1}}e^{i\xi _{1}\ln \frac{S_{t}}{S\ast }%
-(T_{1}-t)\psi (\xi _{1})}\frac{1}{i\xi _{1}}d\xi _{1}\medskip +$\newline
$+2\pi i[\frac{S_{t}e^{-r(T_{2}-t)}}{(2\pi i)^{2}}\int\nolimits_{-\infty
-iw_{2}\omega _{2}}^{+\infty -iw_{2}\omega _{2}}e^{i\xi _{2}\ln \frac{S_{t}}{%
K_{2}}-\Psi (t,0,\xi _{2}-i)}\frac{1}{\xi _{2}}d\xi _{2}-$\newline
$\frac{K_{2}e^{-r(T_{2}-t)}}{(2\pi i)^{2}}\int\nolimits_{-\infty
-iw_{2}\omega _{2}}^{+\infty -iw_{2}\omega _{2}}$ $e^{i\xi _{2}\ln \frac{%
S_{t}}{K_{2}}-\Psi (t,0,\xi _{2})}\frac{1}{\xi _{2}}d\xi _{2}-$ $\frac{%
K_{1}e^{-r(T_{1}-t)}}{2\pi i}e^{-(T_{1}-t)\psi (0)}]=$\newline
$%
F_{2}(S_{t},t;-1,K_{1},T_{1};w_{2},K_{2},T_{2})+F_{1}(S_{t},t;w_{2},K_{2},T_{2})-K_{1}e^{-r(T_{1}-t)} 
$.\bigskip

\noindent In summary, we have verified that the value of a call on an option
equals the value of the corresponding put on the same option plus the value
of the underlying option diminished by $K_{1}e^{-r(T_{1}-t)}$, where $K_{1}$
and $T_{1}$ are the strike price and the maturity date of the compound
option, which is the put-call parity for compound options.\bigskip

6) \textit{Discrete barrier options. }Most analytical pricing formulas for
barrier options assume continuous monitoring of the barrier, which
corrisponds to some practical cases (e.g. in FX markets). However in
practice the barrier might be monitored only at discrete points in time
(e.g., at the close of the market). A discrete barrier option is either
knocked in or knocked out if the price of the underlying asset is across the
barrier at the time it is monitored. In the Gaussian case the pricing
formulas have been studied by [8], [9], [16]; in the L\'{e}vy process models
an interesting survey is presented in [22], where the novel method of [16]
is also discussed. (See also [24] for a numerical approach). In this
subsection we derive a valuation formula for a discrete barrier option as a
further straightforward application of Proposition 2. While there exists
eight barrier options types, depending on the barrier knocking in or out, on
the barrier being above or below the initial value of spot (up or down) and
on the call/put attribute, we confine ourselves to a down-and-out call
without rebate. The other cases can be treated similarly.

Let $B$ denote the level of the barrier and suppose that the underlying
asset is monitored at times $T_{j}$, $j=1,...,M-1$ before the option expiry $%
T_{M}.$ The payoff is $(S_{T_{M}}-K)$\textbf{1}$%
_{M}(S_{T_{j}}>B,j=1,...,M-1;S_{T_{M}}\geq K).$ Therefore Proposition 2 with 
$N=M$ and $A=I$, the $M\times M$ identity matrix, yields:\medskip

$F(S_{t},t)=\frac{Ke^{-r(T_{M}-t)}}{(2\pi i)^{M}}\int\nolimits_{-\infty
-i\omega _{N}}^{+\infty -i\omega _{N}}...\int\nolimits_{-\infty -i\omega
_{1}}^{+\infty -i\omega _{1}}{}^{\text{ }}\frac{1}{(\xi
_{M}+i)\tprod\limits_{k=1,...M-1}\xi _{j}}$\newline
$\ \ \ \ \ \ \ \ \ \ \ \ \ \ \ \ \ \ \ \ \ .\exp
[i\tsum\limits_{j=1}^{M-1}\xi _{j}\ln \frac{S_{t}}{B}+i\xi _{M}\ln \frac{%
S_{t}}{K}-\tsum\limits_{j=1}^{M}(T_{j}-T_{j-1})\psi
(\tsum\limits_{k=j}^{M}\xi _{k})]d\xi _{1}..d\xi _{M}$\newline
with $T_{0}=t$, $\omega _{M}\in ]1,-\lambda _{-}[$ , $\omega _{j}>0$\ and $%
\tsum\limits_{j=1}^{M}\omega _{j}<-\lambda _{-}$.\bigskip

\section{Conclusion}

This work introduces a comprehensive option pricing formula for a very
general family of payoffs, which includes many market-relevant option
payoffs as special cases. The proof is based on Fourier methods and on the
theory of pseudo differential operators that have been successfully applied
in the literature for option pricing in L\'{e}vy models. However, one does
not need to be equipped with such mathematical sophistication in order to
apply the main formula and thus the result may be of interest also for
practitioners. The unifying formula we provide encompasses many existing
option pricing expressions and is a powerful tool for generating new
valuation expressions without effort. We have chosen to focus on discretely
monitored options, as these have received little attention in the
literature, despite their popularity in the trading practice. Each example
is complemented with its Gaussian counterpart and thus, while introducing
new formulas in the general L\'{e}vy setting, the paper may also serve as a
review on discretely monitored options in the traditional Black-Scholes
setting. Finally we stress that the analytical method based on pseudo
differential operators and integration in the complex plane generates a new
numerical method (integration-along-cut method) which often performs better
than the Fast Fourier Transform (see [7]). Therefore numerical computation
will take advantage of our explicit solutions from many a point of view.

\end{document}